# Holographic duality in nonlinear hyperbolic metamaterials

Igor I. Smolyaninov

*Department of Electrical and Computer Engineering, University of Maryland, College Park, MD 20742, USA*

**According to the holographic principle, the description of a volume of space can be thought of as encoded on its boundary. Holographic principle establishes equivalence, or duality, between theoretical description of volume physics, which involves gravity, and the gravity-free field theory, which describes physics on its surface. While generally accepted as a theoretical framework, so far there was no known experimental system which would exhibit explicit holographic duality and be amenable to direct experimental testing. Here we demonstrate that nonlinear optics of hyperbolic metamaterials admits such a dual holographic description. Wave equation which describes propagation of extraordinary light through the volume of metamaterial exhibits 2+1 dimensional Lorentz symmetry. The role of time in the corresponding effective 3D Minkowski spacetime is played by the spatial coordinate aligned with the optical axis of the material. Nonlinear optical Kerr effect bends this spacetime resulting in effective gravitational interaction between extraordinary photons. On the other hand, a holographic dual theory may be formulated on the metamaterial surface, which describes its nonlinear optics via interaction of cylindrical surface plasmons possessing conserved charges proportional to their angular momenta. Potential implications of this duality for superconductivity of hyperbolic metamaterials are discussed.**

Holographic correspondence has been an area of tremendous theoretical activity in recent years. Holographic principle establishes duality between theoretical description of volume physics, which involves gravity, and the gravity-free field theory, which describes physics on its surface. This notion was originally put forward with regard to the relationship between a certain 5D string theory formulated on an anti-de Sitter spacetime (AdS) and a relativistic 4D conformal field theory (CFT) residing on the AdS boundary, so that it is commonly referred to as "AdS/CFT duality" [1]. An extended review of recent progress in holographic principle application to various quantum gravity and high energy physics problems may be found in [2]. More recently, the holographic ideas were further extended to some condensed matter physics situations, including conjectured relevance to superconductivity [3], Fermi liquids [4], graphene [5] and negative index optical metamaterials [6]. However, these extensions were developed by abandoning nearly all the original stringent symmetry conditions of the AdS/CFT duality, and their validity remains unclear. Therefore, solid state physics extensions of holographic principle need considerable theoretical and experimental support. Unfortunately, so far there was no known experimental system which would exhibit unquestionable and explicit holographic duality, and be amenable to direct experimental testing. Here we introduce such a system based on a wire array hyperbolic metamaterial, and demonstrate that its nonlinear optics may be described by either one or another equivalent holographic-dual theories. Both theories are formulated using only conventional, experimentally tested macroscopic electrodynamics and nonlinear optics formalisms. The analog gravity-based "volume theory" is built using the fact that the wave equation which describes propagation of extraordinary light through the volume of metamaterial exhibits 2+1 dimensional Lorentz symmetry. The role of time in the corresponding effective 3D Minkowski spacetime is played by the spatial coordinate aligned with the optical axis of the material. Nonlinear optical Kerr effect bends this spacetime resulting in effective gravitational interaction between extraordinary photons [7]. On the other hand, a holographic dual theory may be formulated on the metamaterial surface, which describes its nonlinear optics via interaction of cylindrical surface plasmons possessing conserved charges proportional to their angular momenta. This theory is built based on a field-theoretical description of nonlinear optics of cylindrical surface plasmons introduced earlier in ref.[8]. Since the latter theory may be extended to encompass mutual interaction of Kaluza-Klein charges, the developed holographic framework may have potential implications for understanding of superconductivity of hyperbolic metamaterials [9]. While our results provide new and important application of holographic principle to a condensed matter system (where there exist just a few known applications), they are also very important for nonlinear optics of metamaterials. They provide new and explicit tool to solve nonlinear Maxwell equations in a difficult situation where magneto-optical coupling appears to be strong by using well-developed tools of effective field theory.

Let us start by outlining the analog gravity-based "volume theory" of nonlinear optics of wire array hyperbolic metamaterials [7]. Light propagation through hyperbolic metamaterials has attracted much recent attention due to their ability to guide and manipulate electromagnetic fields on a spatial scale much smaller than the free space wavelength [10-15]. Almost immediately it was realized that nonlinear optical effects may further increase light confinement in hyperbolic metamaterials based on metal nanowires immersed in a Kerr-type dielectric host (Fig.1(a)) via formation of spatial solitons [16-18]. An interesting counter-intuitive feature of these solitons is that they occur only if a self-defocusing Kerr medium is used as a dielectric host. It appears that this behavior finds natural explanation in terms of analog gravity. Moreover, if gravitational self-interaction is strong enough, a spatial soliton may collapse into a black hole analog with ~50 nm diameter at ~100W laser power [7]. The analog gravity explanation of these effects is based on hyperbolic metamaterial description using effective 2+1 dimensional Minkowski spacetime [19,20].

Let us demonstrate that the wave equation describing propagation of monochromatic extraordinary light inside a hyperbolic metamaterial exhibits 2+1 dimensional Lorentz symmetry. The wire array metamaterial in Fig.1(a) is uniaxial and non-magnetic ($\mu$=1), so that electromagnetic field inside the metamaterial may be separated into ordinary and extraordinary waves (for readers unfamiliar with this common crystal optics terminology let us remind that vector ***E*** of the extraordinary light wave is parallel to the plane defined by the *k*–vector of the wave and the optical axis of the metamaterial). Since hyperbolic metamaterials exhibit strong temporal dispersion, we will work in the frequency domain around some frequency $\omega=\omega_0$. We will assume that in this frequency band the metamaterial may be described by anisotropic dielectric tensor having the diagonal components $\varepsilon_{xx}=\varepsilon_{yy}=\varepsilon_1>0$ and $\varepsilon_{zz}=\varepsilon_2<0$. In the linear optics approximation all the non-diagonal components are assumed to be zero. Propagation of extraordinary light in such a metamaterial may be described by a coordinate-dependent wave function $\varphi_\omega=E_z$ obeying the following wave equation [20]:

$$-\frac{\omega^2}{c^2}\varphi_\omega = \frac{\partial^2 \varphi_\omega}{\varepsilon_1 \partial z^2} + \frac{1}{\varepsilon_2}\left(\frac{\partial^2 \varphi_\omega}{\partial x^2} + \frac{\partial^2 \varphi_\omega}{\partial y^2}\right) \qquad (1)$$

This wave equation coincides with the Klein-Gordon equation for a massive scalar field $\varphi_\omega$ in 3D Minkowski spacetime:

$$-\frac{\partial^2 \varphi_\omega}{\varepsilon_1 \partial z^2} + \frac{1}{(-\varepsilon_2)}\left(\frac{\partial^2 \varphi_\omega}{\partial x^2} + \frac{\partial^2 \varphi_\omega}{\partial y^2}\right) = \frac{\omega_0^2}{c^2}\varphi_\omega = \frac{m^{*2}c^2}{\hbar^2}\varphi_\omega \qquad (2)$$

in which spatial coordinate $z=\tau$ behaves as a "timelike" variable. Eq.(2) describes world lines of massive particles which propagate in a flat 2+1 dimensional Minkowski spacetime [19,20]. The components of metamaterial dielectric tensor define the effective metric $g_{ik}$ of this spacetime: $g_{00}=-\varepsilon_1$ and $g_{11}=g_{22}=-\varepsilon_2$. Similar to our own Minkowski spacetime, the effective Lorentz transformations in the *xz* and *yz* planes form the Poincare group together with translations along *x*, *y*, and *z* axis, and rotations in the *xy* plane. We should also point out that Lorentz symmetry is generally accepted to be broken at the Planck scale – see for example ref.[21] and references therein. This means that similar to hyperbolic metamaterials, metric coefficients of physical vacuum exhibit temporal and spatial dispersion (metric coefficient dependence on the energy-momentum). Mathematically, this dispersion must be expressed as frequency-dependent complex-valued metric coefficients (due to Kramers-Kronig relationship). One cannot guarantee that the wave equation in hyperbolic metamaterials always coincides with Klein-Gordon equation for physical vacuum at the Planck scale, since the latter is yet unknown. However, qualitative similarity of both equations in the case of hyperbolic wire medium has been illustrated in ref.[21].

When the nonlinear optical effects become important, they are described in terms of various order nonlinear susceptibilities $\chi^{(n)}$ of the metamaterial:

$$D_i = \chi_{ij}^{(1)} E_j + \chi_{ijl}^{(2)} E_j E_l + \chi_{ijlm}^{(3)} E_j E_l E_m + ... \qquad (3)$$

Taking into account these nonlinear terms, the dielectric tensor of the metamaterial (which defines its effective metric) may be written as

$$\varepsilon_{ij} = \chi_{ij}^{(1)} + \chi_{ijl}^{(2)} E_l + \chi_{ijlm}^{(3)} E_l E_m + ... \qquad (4)$$

Eq.(4) provides coupling between the matter content (extraordinary photons) and the effective metric of the metamaterial "spacetime". Let us find what kind of simplifications of this general framework may lead to a metamaterial model of usual gravity.

In the weak gravitational field limit the Einstein equation

$$R_i^k = \frac{8\pi\gamma}{c^4}\left(T_i^k - \frac{1}{2}\delta_i^k T\right) \qquad (5)$$

is reduced to

$$R_{00} = \frac{1}{c^2}\Delta\phi = \frac{1}{2}\Delta g_{00} = \frac{8\pi\gamma}{c^4}T_{00} \quad , \tag{6}$$

where $\phi$ is the gravitational potential [22]. Since in our effective Minkowski spacetime $g_{00}$ is identified with $-\varepsilon_1$, comparison of eqs. (4) and (6) indicates that all the second order nonlinear susceptibilities $\chi^{(2)}_{ijl}$ of the metamaterial must be equal to zero, while the third order terms may provide correct coupling between the effective metric and the energy-momentum tensor. These terms are associated with the optical Kerr effect.

Indeed, detailed analysis performed in [7] indicates that Kerr effect in a hyperbolic metamaterial leads to effective gravity. Since $z$ coordinate plays the role of time, while $g_{00}$ is identified with $-\varepsilon_1$, eq.(6) must be translated as

$$-\Delta^{(2)}\varepsilon_1 = \frac{16\pi\gamma^*}{c^4}T_{zz} = \frac{16\pi\gamma^*}{c^4}\sigma_{zz} \quad , \tag{7}$$

where $\Delta^{(2)}$ is the 2D Laplacian operating in the $xy$ plane, $\gamma^*$ is the effective "gravitational constant", and $\sigma_{zz}$ is the $zz$ component of the Maxwell stress tensor of the electromagnetic field in the medium:

$$\sigma_{zz} = \frac{1}{4\pi}\left(D_z E_z + H_z B_z - \frac{1}{2}\left(\vec{D}\vec{E} + \vec{H}\vec{B}\right)\right) \tag{8}$$

A contribution to $\sigma_{zz}$, which is made by a single extraordinary plane wave propagating inside the metamaterial, may be found by assuming without a loss of generality that the $B$ field of the wave is oriented along $y$ direction, so that the other field components may be found from the Maxwell equations as

$$k_z B_y = \frac{\omega}{c}\varepsilon_1 E_x, \quad \text{and} \quad k_x B_y = -\frac{\omega}{c}\varepsilon_2 E_z \tag{9}$$

Taking into account the dispersion law of the extraordinary wave [23]

$$\frac{\omega^2}{c^2} = \frac{k_z^2}{\varepsilon_1} + \frac{k_x^2 + k_y^2}{\varepsilon_2}, \tag{10}$$

the contribution to $\sigma_{zz}$ from a single plane wave appears to be

$$\sigma_{zz} = -\frac{c^2 B^2 k_z^2}{4\pi\omega^2 \varepsilon_1} \tag{11}$$

Thus, for a single plane wave eq.(7) may be rewritten as

$$-\Delta^{(2)}\varepsilon_1 = -\Delta^{(2)}\left(\varepsilon_1^{(0)} + \delta\varepsilon_1\right) = k_x^2 \delta\varepsilon_1 = -\frac{4\gamma * B^2 k_z^2}{c^2\omega^2\varepsilon_1}, \tag{12}$$

where we have assumed that nonlinear corrections to $\varepsilon_1$ are small, so that we can separate $\varepsilon_1$ into the constant background value $\varepsilon_1^{(0)}$ and weak nonlinear corrections. These nonlinear corrections do indeed look like the Kerr effect, assuming that the extraordinary photon wave vector components are large compared to $\omega/c$:

$$\delta\varepsilon_1 = -\frac{4\gamma * B^2 k_z^2}{c^2\omega^2\varepsilon_1 k_x^2} \approx \frac{4\gamma * B^2}{c^2\omega^2\varepsilon_2} = \chi^{(3)} B^2 \tag{13}$$

The latter assumption has to be the case indeed if extraordinary photons may be considered as classic "particles". Eq.(13) establishes connection between the effective gravitational constant $\gamma*$ and the third order nonlinear susceptibility $\chi^{(3)}$ of the hyperbolic metamaterial. Since $\varepsilon_{xx} = \varepsilon_{yy} = \varepsilon_1 > 0$ and $\varepsilon_{zz} = \varepsilon_2 < 0$, the sign of $\chi^{(3)}$ must be negative for the effective gravity to be attractive. For a metal wire array metamaterial shown in Fig.1(a) the diagonal components of the dielectric tensor may be obtained using Maxwell-Garnett approximation [23]:

$$\varepsilon_2 = \varepsilon_z = n\varepsilon_m + (1-n)\varepsilon_d \tag{14}$$

$$\varepsilon_1 = \varepsilon_{x,y} = \varepsilon_d \frac{\varepsilon_m(1+n) + \varepsilon_d(1-n)}{\varepsilon_d(1+n) + \varepsilon_m(1-n)} \qquad (15)$$

where $n$ is the volume fraction of the metallic phase, and $\varepsilon_m$ and $\varepsilon_d$ are the dielectric permittivities of the metal and dielectric phase, respectively. For a typical metal $-\varepsilon_m \gg \varepsilon_d$, and at small $n$ eq.(15) may be simplified as

$$\varepsilon_1 \approx \varepsilon_d \frac{(1+n)}{(1-n)} \sim \varepsilon_d \qquad (16)$$

Thus, in order to obtain attractive effective gravity the dielectric host medium must indeed exhibit negative (self-defocusing) Kerr effect. Extraordinary light rays in such a medium will behave as 2+1 dimensional world lines of self-gravitating bodies and may collapse into sub-wavelength spatial solitons.

As a next step, let us follow the general spirit of AdS/CFT duality and arrange for an effective "gravitational horizon" at some radius $\rho$ from the metamaterial sample center. A general recipe for such an analog horizon has been described in [24]. Near horizon its surface may be considered as almost flat so that a case of constant $\varepsilon_2 = \varepsilon_z < 0$ and finite $\varepsilon_1(x) = \varepsilon_x = \varepsilon_y$ which changes sign from $\varepsilon_1 > 0$ to $\varepsilon_1 < 0$ as a function of $x$ in some frequency range around $\omega = \omega_0$ may be assumed. Because of translational symmetry along the $z$ direction, we may consider a plane wave solution in the $z$ direction with a wave vector component $k_z$. Introducing $\psi = B$ as above (see eq.(9)), we obtain wave equation

$$-\frac{\partial^2 \psi}{\partial x^2} + \frac{\varepsilon_2 k_z^2}{\varepsilon_1}\psi = \frac{\varepsilon_2 \omega_0^2}{c^2}\psi \ , \qquad (17)$$

where $V=\varepsilon_2/\varepsilon_1$ plays the role of effective potential. Its analysis indicates that $E_x \sim -\frac{1}{\varepsilon_x}\frac{\partial \psi}{\partial z}$ diverges at the horizon and the choice of $\varepsilon_1 = \alpha x^2$ (where $\alpha > 0$) leads to Rindler-like optical space near $x=0$ [24]. On the other hand, as has been shown in ref.[25], it is also relatively easy to emulate an AdS "cosmological horizon" by gradual variation of volume fraction $n$ of the metallic phase in eqs.(14,15). In both cases a gradual transition from $\varepsilon_{xx} = \varepsilon_{yy} = \varepsilon_1 > 0$ to $\varepsilon_1 < 0$ must occur at $r=\rho$, while $\varepsilon_{zz} = \varepsilon_2 < 0$ must remain negative. In the metal wire array geometry shown in Fig.1(a) these conditions may be satisfied around

$$n \approx \frac{\varepsilon_m + \varepsilon_d}{\varepsilon_m - \varepsilon_d} \qquad (18)$$

by gradual increase of volume fraction $n$ of metallic nanowires. As a result, as shown in Fig.1(b), a hyperbolic metamaterial described by 2+1 dimensional effective Minkowski spacetime will exist inside the $r=\rho$ cylinder, while the metamaterial outside the cylinder surface will behave as anisotropic metal. Such an experimental geometry may be considered as a metamaterial waveguide, which supports electromagnetic modes having divergent electric field $E_x$ on its surface. The field decays exponentially into the anisotropic metal outside the waveguide. Such guided modes are usually called cylindrical surface plasmons (CSP), which "live" at the $r=\rho$ interface [8]. Let us demonstrate that nonlinear optics of these CSPs may be formulated as an effective field theory, which is holographic dual to the effective 2+1 gravity described above. Such a theory may be formulated similar to the field-theoretical description of nonlinear optics of CSPs living on surfaces of single nanoholes and nanowires [8], which has strong similarity with Kaluza-Klein theories.

Similarity between the nonlinear optics of CSPs and the Kaluza-Klein theories stems from the way in which electric charges are introduced in the original five-dimensional Kaluza-Klein theory (see for example [26]). In this theory the electric charges are introduced as chiral (nonzero angular momentum *L*) modes of a massless quantum field, which is quantized over the cyclic compactified fifth dimension. In a similar fashion, nonlinear optics of CSPs may be formulated as a field theory in a curved 2+1 dimensional space-time defined by the metal interface which has an extended z-coordinate and a small "compactified" angular $\phi$ dimension along the circumference of the cylinder. The resulting 1+1 dimensional effective field theory of CSP interaction describes higher order (*L>0*) CSP modes as having quantized effective chiral charges equal to their angular momenta *L*. These massive slow moving effective charges exhibit long-range interaction via exchange of fast massless CSPs having zero angular momentum.

In order to look similar to Kaluza-Klein theory, the medium should be either optically active, or exhibit magnetic field induced optical activity. Such a medium would be able to discriminate between CSP waves which have opposite angular momenta, and thus expected to have opposite effective Kaluza-Klein charges. The best way to accommodate this requirement is to consider zero angular momentum CSP modes as quanta of the "gyration field" (the field of the gyration vector *g*), which relates the *D* and *E* fields in an optically active medium [27]:

$$\vec{D} = \vec{\vec{\varepsilon}}\vec{E} + i\vec{E} \times \vec{g} \qquad (19)$$

If the medium exhibits magneto-optical effect, and does not exhibit natural optical activity, *g* is proportional to the magnetic field *H*:

$$\vec{g} = f\vec{H}, \qquad (20)$$

where the constant $f$ may be either positive or negative. For metals in the Drude model at $\omega >> eH/mc$ the magnetic field induced optical activity is defined by

$$f(\omega) = -\frac{4\pi N e^3}{cm^2\omega^3} = -\frac{e\omega_p^2}{mc\omega^3}, \qquad (21)$$

where $\omega_p$ is the plasma frequency and $m$ is the electron mass [27]. Comparison of eqs.(3,19,20) indicates that introduction of gyration field constitutes an alternative way of treating third order nonlinear optical effects responsible for the effective gravity of the bulk theory. Moreover, it is easy to demonstrate that so introduced gyration field would indeed lead to the Coulomb-like interaction of effective "chiral charges".

Let us consider Maxwell equations in the presence of the axial gyration field $g_\phi = g_\phi(r,z,t)$. In order to illustrate essential physics, let us consider solutions at $r>\rho$ where the metamaterial may be treated as anisotropic metal, and neglect metal anisotropy for the sake of simplicity. After simple calculations the wave equation in such a case may be written in the form

$$-\Delta \vec{B} = -\frac{\varepsilon}{c^2}\frac{\partial^2 \vec{B}}{\partial t^2} + \frac{i}{c}\frac{\partial(\vec{\nabla}\times[\vec{E}\times\vec{g}])}{\partial t} \qquad (22)$$

which for the z-components of a solution proportional to $\sim e^{iL\phi}$ may be re-written as

$$\frac{\partial}{r\partial r}\left(r\frac{\partial B_z}{\partial r}\right) + \frac{\partial^2 B_z}{\partial z^2} - \frac{\varepsilon\partial^2 B_z}{c^2\partial t^2} - \frac{L^2}{r^2}B_z + \frac{iLg}{rc}\frac{\partial(iE_z)}{\partial t} + \frac{iL}{cr}\left(\frac{\partial g}{\partial t}\right)(iE_z) = 0 \qquad (23)$$

The latter equation is similar to the Klein-Gordon equation, in which $L$ and $g/r$ play the role of the effective charge, and the effective potential, respectively [8] (here it is important to mention that for such higher order CSP modes $iE_z \sim B_z$, where the coefficient of proportionality is determined by the boundary conditions). Thus, action of gyration field $g$ on chiral charges is similar to action of electric field on electric charges.

On the other hand, we may solve the nonlinear Maxwell equations and explicitly demonstrate that the higher angular momentum modes (the chiral charges) behave as the sources of gyration field (the field of the fundamental $L=0$ CSP mode). Let us search for the solutions of the nonlinear Maxwell equation (22) of the form **B**=**B**$_0$+**B**$_L$ and **E**=**E**$_0$+**E**$_L$, where **B**$_0$ and **B**$_L$ are the fundamental mode, and the $L>0$ guided mode, respectively. The gyration field may be obtained in a self-consistent manner as **g**=f(**B**$_0$+**B**$_L$). We are interested in the solution for the field **B**$_0$ in the limit of small frequencies $\omega_0$ in the presence of **B**$_L$ field, so that in the found solution the **B**$_L$ field will act as a source of **B**$_0$. After neglecting the terms proportional to $f^2$ in eq.(22), and taking into account that **B**$_0$ and **B**$_L$ are incoherent solutions of linear Maxwell equations, we obtain

$$\Delta \vec{B}_0 = \frac{4\pi \omega_L f}{c^2} \vec{\nabla} \times \vec{S}_L \qquad (24)$$

where $S_L$ is the Pointing vector of the $L$-th mode. This equation is similar to the Poisson equation in which the term $\vec{\nabla} \times \vec{S}_L$ acts as a source. Moreover, using vector calculus we may also derive an analog of the Gauss theorem for the effective chiral charges. Let us consider a cylindrical volume $V$ around the "metamaterial waveguide" formed by the effective horizon at $r=\rho$ (see Fig.2), such that the side wall of the volume $V$ is located very far from the waveguide and the electromagnetic field is zero at this wall. If $S$ is the closed two-dimensional cylindrical surface bounding $V$, with area element $da$ and unit outward normal **n** at $da$, and $S_1$ and $S_2$ are the front and the back surfaces of $V$, we may write the following integral equation for the Pointing vector $S_L$ of the $L$-th mode:

$$\int_V \vec{\nabla} \times \vec{S}_L d^3x = \int_{S2} \vec{N} \times \vec{S}_L da - \int_{S1} \vec{N} \times \vec{S}_L da \qquad (25)$$

where *N* is the chosen direction of the metamaterial waveguide. Using equation (24) we obtain

$$\int_V \frac{4\pi f \omega_L}{c^2} \vec{\nabla} \times \vec{S}_L d^3 x = \int_{S2} \vec{N} \times [\vec{\nabla} \times \vec{B}_0] da - \int_{S1} \vec{N} \times [\vec{\nabla} \times \vec{B}_0] da \qquad (26)$$

Since $\vec{N} \times [\vec{\nabla} \times \vec{B}_0] = \frac{\partial B_{0\phi}}{\partial z}$, we see that a "chiral charge" produces a local step in the gyration field. Eqs. (24) and (26) (which represent effective Poisson equation, and effective Gauss Theorem for chiral charges, respectively) clearly demonstrate that the "chiral charges" interact according to the one-dimensional Coulomb law with the interaction energy growing linearly with distance (in reality this idealized linear growth is cut off by the absorption in the metamaterial). This field-theoretical description of nonlinear optics of chiral charges provides a holographic dual to the effective gravity description discussed earlier. The described duality appears useful for exactly the same reason as the original AdS/CFT duality is useful in high energy physics. Nonlinear optical Maxwell equations are very difficult to study and analyze either numerically or analytically. This task is especially difficult in the case of newly developed sophisticated metamaterials, where the usual experience does not always apply. For example, increased light confinement in hyperbolic metamaterials due to formation of spatial solitons [16-18] is counter-intuitive. It occurs only if a self-defocusing Kerr medium is used as a dielectric host. On the other hand, this behavior finds simple and natural explanation in terms of analog gravity [7]. Similar to the original AdS/CFT correspondence, the duality between effective gravity and effective field theory of chiral charges described above provides us with a new powerful tool to understand nonlinear optics of metamaterials when the nonlinear interactions are strong. Moreover, similar to AdS/CFT, these dual descriptions work best in the opposite limits: strong chiral interaction in 1+1D space corresponds to weak gravity limit in 2+1D, and vice versa.

It is also interesting to note that due to similar topological origin, theoretical description based on chiral charges may be extended to encompass mutual interaction of Kaluza-Klein electric charges [8]. This approach leads to a picture of strong plasmon-mediated electron-electron interaction in metal nanowire and metal nanohole arrays. Therefore, the developed holographic framework may have potential implications for understanding of superconductivity in metal wire array hyperbolic metamaterials. It appears that similar to a different metamaterial approach discussed in ref.[9], conditions favorable for strong electron-electron interaction arise in the epsilon near zero (ENZ) regime, which corresponds to formation of an effective "gravitational horizon" (see Fig.1(b)). This feature of our model is somewhat similar to the holographic approach to superconductivity developed in [3]. Indeed, within the scope of holographic models the superconducting transition is typically linked to classical instability of a black hole horizon in anti-de Sitter (AdS) space against perturbations by a charged scalar field. The instability appears when the black hole has Hawking temperature $T = T_c$ [3].

Another interesting application of the developed holographic dual formalisms appears to be consideration of the superconducting state of physical vacuum in a strong magnetic field. As demonstrated by Chernodub [28], strong magnetic field forces vacuum to develop real condensates of electrically charged $\rho$ mesons, which form an anisotropic inhomogeneous superconducting state similar to Abrikosov vortex lattice. As far as electromagnetic field behavior is concerned, this state of vacuum constitutes a hyperbolic metamaterial [29]. Moreover, it was demonstrated in ref.[21] that a well-known "additional wave" solution of macroscopic Maxwell equations describing metamaterial optics of vacuum in the presence of dispersion leads to prediction of ~ 2 GeV "heavy photons" in vacuum subjected to a strong magnetic field [2]. These "heavy photons" may be identified with cylindrical surface plasmons propagating along individual Abrikosov vortices. Holographic consideration presented above leads to conclusion that heavy photon states which have non-zero angular momenta must behave as interacting "chiral charges". The chiral charge of a heavy photon is equal to its angular momentum *L*. These massive slow moving effective charges exhibit long-range interaction via exchange of fast massless CSPs having zero angular momentum. Effective field theory of these chiral charges may be formulated similar to the effective field theoretical description of nonlinear optics of wire array hyperbolic metamaterials presented above. Thus, our dual holographic description is directly applicable to real physical vacuum subjected to a strong magnetic field.

In summary, we have demonstrated that nonlinear optics of hyperbolic metamaterials admits explicit dual holographic descriptions both in terms of 2+1 dimensional bulk effective gravity, and in terms of 1+1 dimensional surface effective field theory. To our knowledge, this is the first known experimental system which exhibits such an explicit holographic duality and is amenable to direct experimental testing. Such testing may be performed using ferrofluid-based self-assembled hyperbolic metamaterials [30], which exhibit considerable negative Kerr effect (and therefore effective attractive gravity-like interaction [7]) due to thermal expansion of kerosene in the ferrofluid. Gradual variation of the dielectric tensor components $\varepsilon_{xx}= \varepsilon_{yy}$ and $\varepsilon_{zz}$ of the ferrofluid may be achieved due to gradient of external magnetic field and/or self-focusing, leading to appearance of an effective horizon. On the other hand, illumination of ferrofluid with light having non-zero orbital angular momentum [31] will lead to excitation of cylindrical surface plasmons having non-zero angular momentum propagating inside the ferrofluid. As demonstrated by eqs.(23) and (26), such cylindrical surface plasmons will exhibit strong 1D Coulomb-like interaction, which will lead to nonlinear dependence of ferrofluid transmission on illuminating power, similar to nonlinear effects in plasmon-mediated nonlinear optical transmission through individual nanoholes [32] and nanohole arrays [33].

**Figure Captions**

**Figure 1.** (a) Typical geometry of metal ($\varepsilon_m$) wire array nonlinear hyperbolic metamaterial. The dielectric host ($\varepsilon_d$) exhibits nonlinear optical Kerr effect. (b) Gradual variation of volume fraction $n$ of metal wires (shown from the top) leads to appearance of an effective horizon at $r=\rho$. Such an experimental geometry supports cylindrical surface polaritons (CSP) at the $r=\rho$ interface.

**Figure 2**. Schematic view of the metamaterial waveguide formed by the effective horizon at $r=\rho$ from Fig.1(b), and the front and back surfaces $S_1$ and $S_2$ of the auxiliary cylinder used in the derivation of the Gauss theorem for cylindrical surface plasmons treated as "chiral charges" (see eq.(25)). Vector $N$ indicates chosen direction of the metamaterial waveguide.

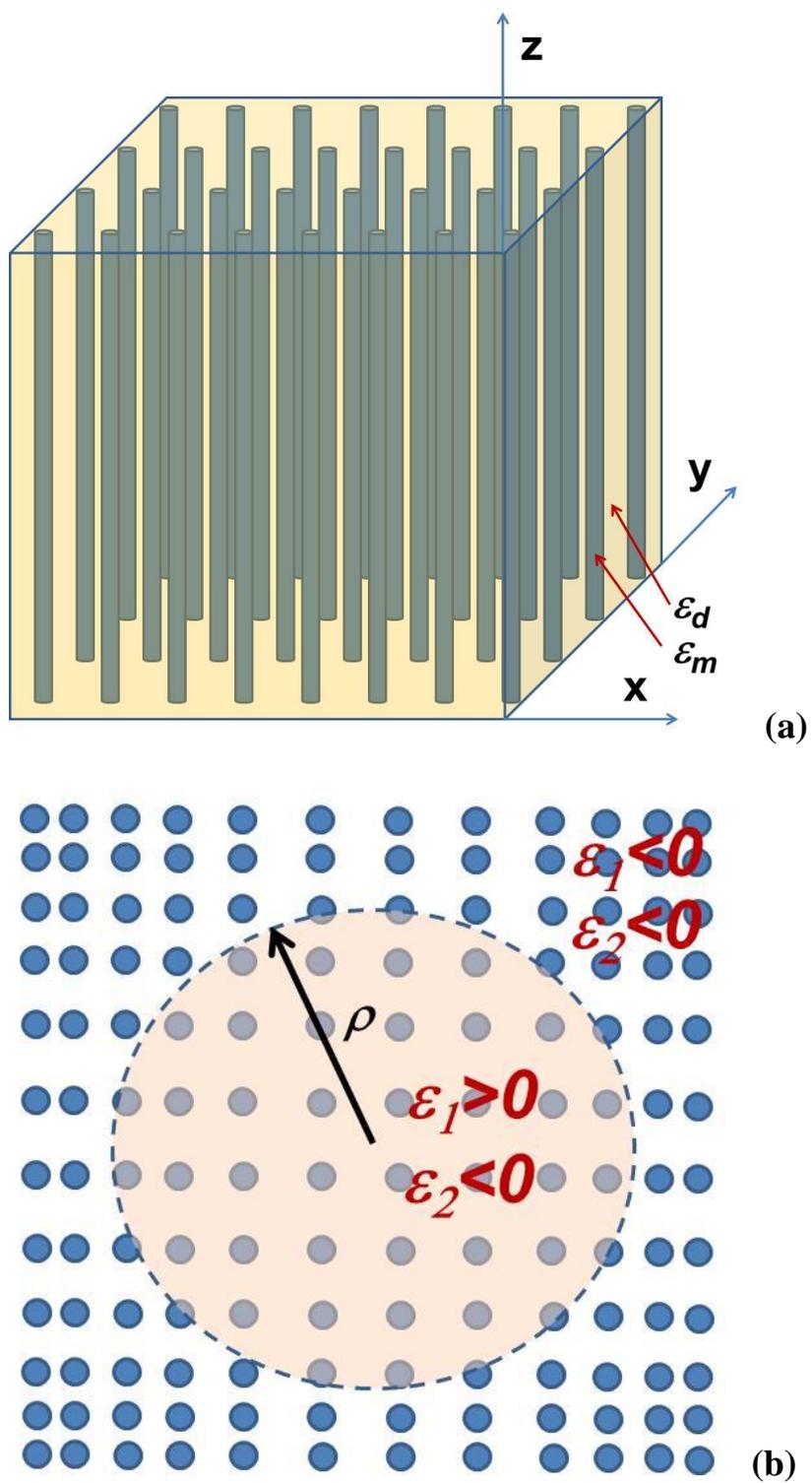

**Fig. 1**

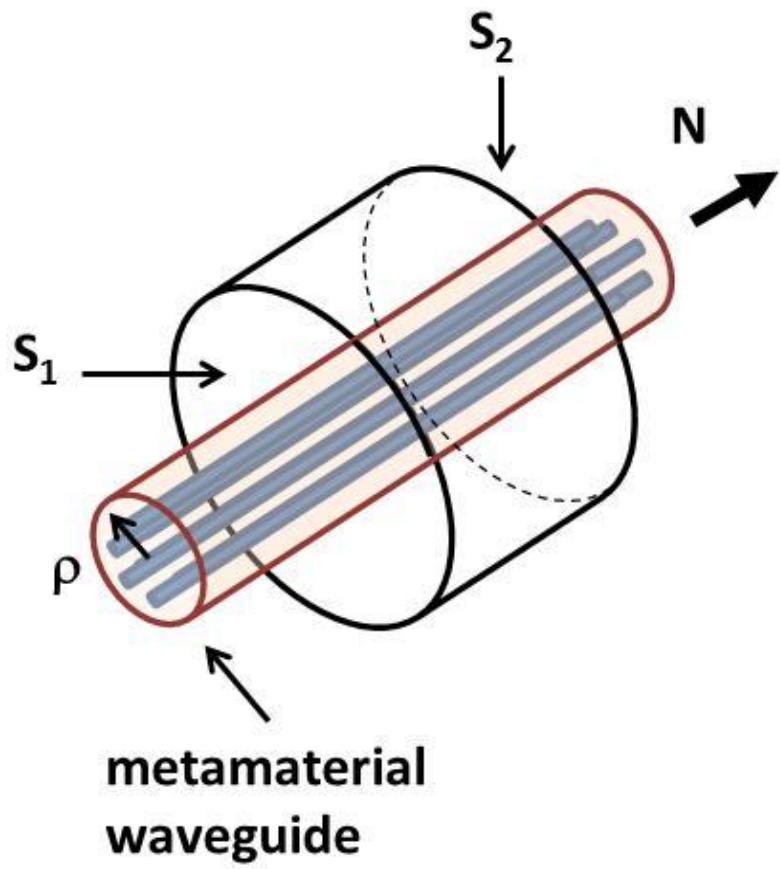

**Fig. 2**